\documentclass[prb,floatfix,amsfonts,twocolumn]{revtex4}
\usepackage{graphicx,graphics,color,epsfig}
\usepackage{bm}
\usepackage{amsmath}
\usepackage{amssymb}

\newcommand{\bQ}{{\bf Q}}
\newcommand{\bG}{{\bf G}}
\newcommand{\bk}{{\bf k}}

\newcommand{\br}{{\bf r}}

\newcommand{\bq}{{\bf q}}

\newcommand{\beqa}{\begin{eqnarray}}
\newcommand{\eeqa}{\end{eqnarray}}

\begin{document}
\preprint{}
\title{Wigner Supersolid of Excitons in Electron-hole Bilayers}
\author{Yogesh N. Joglekar$^1$, Alexander V. Balatsky$^2$, S. Das
Sarma$^3$}
\affiliation{$^1$ Department of Physics, Indiana University - Purdue
University Indianapolis, Indiana 46202 \\ $^2$ Theoretical Division, Los
Alamos National Laboratory, Los Alamos, New Mexico 87545\\ $^3$ Condensed 
Matter Theory Center, Department of Physics, University of Maryland, 
College Park, Maryland 20742}
\date{\today}
\begin{abstract}
Bilayer electron-hole systems, where carriers in one layer are electrons and
carriers in the other are holes, have been actively investigated in recent
years with the focus on Bose-Einstein condensation of excitons. This
condensation is expected to occur when the layer separation $d$ is much
smaller than the interparticle distance $r_sa_B$ within each layer. In this 
note, we argue for the existence of a state, Wigner supersolid, in which 
excitons are phase-coherent but form a Wigner crystal due to dipolar repulsion. 
We present the qualitative phase diagram of bilayer system, and discuss
properties and possible signatures of the Wigner supersolid phase.
\end{abstract}
\maketitle



\noindent{\it Introduction:} The phenomenon of Bose-Einstein 
condensation (BEC) where the many-body wavefunction for the ground
state of a macroscopic number of bosons is described by a uniform
phase and acquires phase rigidity, is a remarkable manifestation of
interplay between quantum mechanics and statistics of the particles.
This phenomenon does not depend on the microscopic structure of
bosons or their interactions, and does not make assumptions about
possible broken symmetries in the ground state. Based on these
observations, Moskalenko, Blatt, and Keldysh predicted that excitons - 
metastable bound states of electron-hole pair - in semiconductors will undergo
BEC under appropriate circumstances.~\cite{keldysh} Electron-hole
bilayers are expected to exhibit a {\it uniform} BEC of excitons -
dipolar superfluid - when the distance between the two layers $d$
is much smaller than the typical interparticle distance $r_s
a_B$.~\cite{lozo,rice,shev} Here $a_B=\hbar^2\epsilon/m^{*}e^2$ is
the Bohr radius for a particle with band mass $m^{*}$ in a
semiconductor with dielectric constant $\epsilon$ (For typical
semiconductors, $\epsilon\sim 13$ and $m^{*}\sim 0.10\, m_e$ implies
that $a_B\sim$ 100 \AA). In recent years, advances in sample
preparations have made it possible to fabricate samples in which the
two layers are close ($d \sim$ 100-300 \AA), the exciton lifetime is
relatively long, and the carrier mobilities at low densities are
high or, equivalently, disorder effects are
small.~\cite{butov,snoke,lai,em,sivan} Therefore, an experimental
exploration of the entire phase-diagram of electron-hole bilayer
system will be feasible in near future.~\cite{recent}

Recent experiments have sparked interest in two disparate aspects of BEC, 
namely its realization in semiconductors and in solid Helium under 
pressure.~\cite{he4} These two aspects, condensation of (metastable) bosons 
in a 
strongly interacting semiconductor environment and non-uniform Bose-Einstein 
condensates, address fundamental questions such as what are necessary 
and sufficient conditions for Bose-Einstein condensation? Is superfluidity 
related to a {\it uniform} Bose-Einstein condensate? {\it In this note, 
we show that electron-hole bilayers support a ground state which 
combines these two features, namely Bose-Einstein condensation and broken 
translational symmetry.} 

In the following, we first discuss zero-temperature phase diagram of 
electron-hole bilayer system. Then we present qualitative description 
of various phase boundaries and focus on the transition from the 
dipolar superfluid phase to the Wigner supersolid phase. We end the note 
with discussion regarding experimental signatures of the Wigner supersolid 
phase and conclusions. 

\begin{figure}[thbp]
\begin{center}
\includegraphics[width=3.2in]{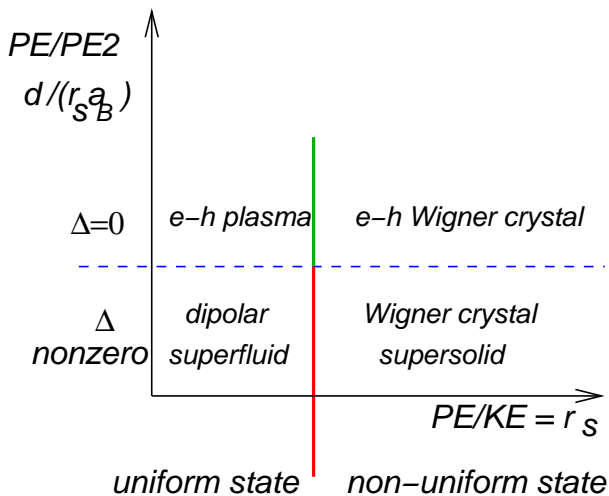}
\caption{Schematic phase diagram of a system with two dimensionless 
parameters $d/(r_sa_B)$ and $r_s$, showing four possible phases. We emphasize 
that the details of the topology, for example, the dependence of Wigner 
crystallization threshold $r_s$ on the ratio of intralayer-to-interlayer 
potential energy $d/(r_sa_B)$, are not known. Therefore, the intersection of 
the two lines should not be considered a multi-critical point. The four 
possible ground states are i) uniform dipolar superfluid ii) electron-hole 
plasma iii) uncorrelated Wigner crystals, and iv) phase-coherent bilayers 
with broken translational symmetry. The first three have been extensively 
discussed in the literature.}
\label{fig: phases}
\end{center}
\vspace{-5mm}
\end{figure}

\noindent{\it Phase Diagram:} 
Let us consider a bilayer system with electrons in the top layer and 
holes in the bottom layer. We choose a convention such $e^\dagger(\br)$ 
denotes an operator which creates electron at position $\br$ and 
$h^{\dagger}(\br)$ denotes an operator which creates a hole at position 
$\br$ in the bottom layer. The density matrix of this system has four 
components. $\rho_{ee}(\br,\br')=\langle e^{\dagger}(\br)e(\br')\rangle$ and 
$\rho_{hh}(\br,\br')=\langle h^{\dagger}(\br)h(\br')\rangle$ denote density 
matrices for electrons and holes, and $\Delta(\br,\br')=\langle e^{\dagger}
(\br)h^{\dagger}(\br')\rangle=\Delta(\br',\br)^*$ denotes the interlayer 
phase-coherence matrix elements. This system is characterized by two 
dimensionless parameters which can be tuned independently. The first, 
$d/(r_sa_B)$, is the ratio of intralayer and interlayer Coulomb interactions 
(PE/PE2), and is a measure of ``phase coherence'' between the two layers. This 
parameter, for $d/(r_sa_B)<1$, drives the quantum phase transition from 
uncorrelated bilayers, $\Delta=0$, to a state with phase-coherent bilayers, 
$\Delta\neq 0$. We emphasize that here we have assumed the simplest scenario, 
in which the formation of individual excitons and the establishment of 
long-range phase coherence happen simultaneously. In a more general case, 
there will be two phase transitions, one corresponding to each of the two 
possibilities outlined above. The second parameter $r_s$ is the ratio of 
potential energy $e^2/\epsilon(r_s a_B)$ and kinetic energy 
$\hbar^2/m^*(r_sa_B)^2$ of carriers within a single layer (PE/KE). This 
parameter, for $r_s\gg 1$, drives the
quantum phase transition from a uniform density state,
$\rho_{ee}(\bq)=\rho_{hh}(\bq)\propto\delta_{\bq,0}$, to the Wigner crystal 
state with broken translational symmetry, $\rho_{ee}(\bq)=\rho_{hh}(\bq)
\propto\delta_{\bq,{\bf G}}$ where $\bG$ is a reciprocal lattice vector for 
Wigner crystal. Therefore, in the simplest scenario, we 
expect $2^2=4$ possible distinct ground states (Figure~\ref{fig: phases}). 
For small $d/(r_sa_B)$ and small $r_s$ ($d/a_B\le r_s\le 1$), the ground state 
of the system is a uniform Bose-Einstein condensate of excitons or a dipolar
superfluid.~\cite{palo,dipolar} At large $d/(r_sa_B)$ and small 
$r_s$, the ground state is a uniform 2-component (electron-hole) plasma. At
large values of $d/(r_s a_B)$ and large $r_s$ ($d/a_B\gg r_s\gg 1$) the
system consists of Wigner crystals in respective layers which are weakly
correlated. These three phases have been considered in the 
literature.~\cite{palo} In particular, excitonic superfluid and Wigner 
crystal, being broken symmetry states, have been extensively studied. For 
example, spontaneous interlayer coherence has been observed in quantum Hall 
bilayers near total filling factor one~\cite{em} although 
compelling evidence for true off-diagonal long-range superfluid order is 
still somewhat ambiguous. Similarly, it is widely believed~\cite{tanatar} 
that a single electron (or hole) layer becomes a Wigner crystal for large 
$r_s=r_{s0}\sim 40$ although the experimental evidence is not conclusive.

In this note, we focus on the fourth possible ground state, which 
{\it will} occur when $d/(r_sa_B)\leq 1$ and $r_s\gg 1$. We point out that 
the two broken-symmetry states - dipolar excitonic condensate and Wigner 
crystal - will coexist in this regime. Based on general principles, this phase
will have a ground state with phase-coherence between the two layers and a
broken translational symmetry: $\Delta\neq 0$ and $\rho_{ee}(\bq)=
\rho_{hh}(\bq)\propto\delta_{\bq,\bQ}$.{\it We propose that this state
is a Wigner crystal of phase-coherent excitons: a Wigner supersolid.}

\begin{figure}[thbp]
\begin{center}
\includegraphics[width=3.2in]{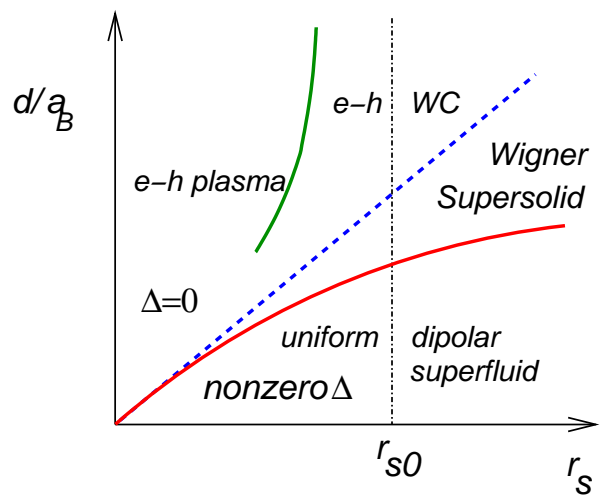}
\caption{Schematic phase diagram of the electron-hole bilayer. Standard
phases discussed in the literature are electron-hole Wigner crystal 
($d/a_B\gg 1$, $r_s\gg 1$), 2-component plasma ($d/a_B\gg 1$, $r_s\leq 1$) and 
dipolar superfluid ($d/a_B\leq 1$, $r_s\gg 1$). We postulate that a Wigner 
crystal of phase-coherent excitons exists in the region 
$\sqrt{r_s}\leq d/a_B\leq r_s$, between the dipolar superfluid and the 
uncorrelated Wigner crystals. We call this phase Wigner supersolid (WS) 
because it shows broken translational symmetry in the diagonal density matrix 
elements, $\rho_{ee}(\bq)=\rho_{hh}(\bq)\propto\delta_{\bq\bG}$, as well as 
interlayer phase coherence, $\Delta\neq 0$, in the off-diagonal density 
matrix elements. The transition between dipolar superfluid and WS is of 
first order as any liquid-to-solid transition; on the other hand, the 
transition between WS and electron-hole Wigner crystal is of second order 
as it is associated with the disappearance of phase coherence.}
\label{fig: newphases}
\end{center}
\vspace{-8mm}
\end{figure}

Now, we motivate the existence of this phase and show how various phase 
boundaries can be qualitatively understood. Figure~\ref{fig: newphases} shows 
the (same) phase diagram of a bilayer electron-hole system~\cite{palo} as a 
function of $d/a_B$ and $r_s$ (solid lines only). We choose these variables, 
instead of the ratio of energies, as the axes because experimentally these 
two can be tuned independently. This phase diagram does not take into account 
the transition from phase-coherent bilayers ($\Delta\neq 0$) to uncorrelated 
bilayers ($\Delta=0$) which happens at $d/a_B\sim r_s$ (dotted blue line). 
Let us consider the region between the solid (green and red) lines from two 
different limits, increasing $d/a_B$ at moderate $r_s$ and increasing $r_s$ at 
moderate $d/a_B$. First, we start from the uniform dipolar superfluid state 
characterized by uniform phase coherence $\Delta\neq 0$ and uniform density 
$\rho_{ee}(\bq)=\rho_{hh}(\bq)=n_0\delta_{\bq,0}$. The kinetic energy per 
exciton~\cite{cav1} is given by $\hbar^2/m^{*}(r_sa_B)^2$ whereas the 
potential energy due to dipolar repulsion is given by 
$e^2d^2/\epsilon(r_sa_B)^3$. Therefore, when $\sqrt{r_s}\leq d/a_B$ potential 
energy dominates the kinetic energy and the excitons will form a hexagonal 
Wigner crystal~\cite{gabor,palo2} to minimize the potential energy. This 
argument gives the phase boundary between the uniform dipolar superfluid 
state, and the Wigner crystal of excitons characterized by $\Delta\neq 0$ and
$\rho_ee(\bq)=\rho_hh(\bq)\propto\delta_{\bq,\bG}$ (solid red line). 
{\it Such a phase with dipolar exciton condensate and crystalline 
structure within each layer is, by definition, a supersolid}. Since this is a 
transition from a uniform state to a crystalline state, in the absence of 
disorder, it will be a {\em first-order} transition.

The existence of this phase can also be argued if we start with system in
the uniform 2-component plasma state characterized by $\Delta=0$ and 
$\rho_{ee}(\bq)=\rho_{hh}(\bq)=n_0\delta_{\bq,0}$. As $r_s$ increases, each 
layer undergoes hexagonal Wigner crystallization at a critical value of 
$r_s(d/a_B)$, leading to a ground state with no phase coherence, $\Delta=0$, 
but crystalline density modulations, $\rho_{ee}(\bq)=\rho_hh(\bq)\propto 
\delta_{\bq,\bG}$.~\cite{tanatar} At large $d/a_B\gg 1$, this value of $r_s$ 
is roughly independent of the value of $d/a_B$, and will approach the 
single-layer value $r_{s0}\sim 38$ asymptotically (solid green line). The 
behavior of this phase boundary at small $d/a_B>r_s$ will, in general, be 
nontrivial, due to interlayer interactions playing a role in determining the 
ground state. 
Now as $d/a_B$ is reduced or, equivalently, $r_s$ is increased, when 
$d/a_B\leq r_s$ (dotted blue line), the two Wigner crystals will become 
phase-coherent, $\Delta\neq 0$, and will still maintain the broken 
translational symmetry, $\rho_{ee}(\bq)=\rho_{hh}(\bq)\propto
\delta_{\bq,\bG}$, thus forming a hexagonal Wigner crystal of phase-coherent 
excitons.~\cite{gabor,tanatar,palo2} This transition between two layers 
with identical crystal structure become is associated with the appearance of 
phase coherence and is therefore {\it continuous}.

We emphasize that this novel phase, a Wigner supersolid, is possible
only due to specific properties of electron-hole bilayers. Traditionally,
supersolid phase discussed in the literature has been mostly in the context
of $^4$He.~\cite{he4} Some recent work has introduced the possibility of a
supersolid phase in cold-atom optical lattice systems with extended Hubbard
interactions~\cite{scarola} as well as in lattice models of hard-core
bosons with repulsion.~\cite{hcbosons} Since excitons in bilayer systems
have dipolar repulsion, which provides the incentive for localization, a
Bose-Einstein condensate with broken translational symmetry is possible. In
a single layer system, carriers undergo Wigner crystallization but since they
are fermions there is no phase coherence. On the other hand, bulk
semiconductors support excitonic condensation. However, since the exciton
interaction is not necessarily repulsive (due to random orientation of
exciton dipoles), there is no Wigner crystallization. Electron-hole systems,
in which all excitons have the same dipole moment, offer a unique
realization of hard-core bosons with repulsive interactions in a
semiconductor environment.

Now we sketch a simple model for phase transition from the uniform dipolar 
superfluid phase to the Wigner supersolid phase. The dipolar superfluid is 
characterized by a nonzero order parameter $\Delta=|\Delta|\exp[i\Phi]$ 
where $\Phi$ is the dipolar phase. Note that the density of this superfluid 
is uniform. The long-wavelength low-energy dynamics of a dipolar superfluid 
is described by the action~\cite{dipolar}
\begin{equation}
{\cal S}_0=\sum_{\bk,\omega}\left[
\omega\rho_{\bk\omega}\Phi_{\bk\omega}
-\frac{\rho_d}{2}k^2\Phi^{*}_{\bk\omega}\Phi_{\bk\omega}+
\frac{1}{2}S^{-1}_{\bk\omega}\rho^{*}_{\bk\omega}\rho_{\bk\omega}
\right].
\label{EQ:S1}
\end{equation}
Here, $\rho_d$ is the superfluid phase stiffness, $\rho_{\bk\omega}$
is the Fourier transform of the dipolar density fluctuation, and the
dipolar phase and condensate number satisfy $[\rho,\Phi]=i$.  We
introduce $S^{-1}_{\bk\omega}\rho^{*}_{\bk\omega}\rho_{\bk\omega}$
term in the action to account for the dipolar density-density
interaction. At small wavevectors, this interaction gives the
capacitive mass-term for uniform density fluctuations. Integrating
the density fluctuations, we arrive at the action for the dipolar
phase,
\begin{equation}
{\cal S}_\Phi=\frac{1}{2}\sum_{\bk,\omega}(S_{\bk\omega}\omega^2
-\rho_d k^2) \Phi^{*}_{\bk\omega}\Phi_{\bk\omega}.
\end{equation}
The dispersion of the collective mode is given by $\omega_\bk=
k\sqrt{\rho_d/S_{\bk\omega}}$. When $k\rightarrow 0$, this gives the
linearly dispersing sound mode, $\omega_\bk=k\sqrt{\rho_d/C}=v_ck$,
where $C = S_{\bk = 0, \omega = 0}$ is the capacitance. Since the
dipolar fluid is a uniform Bose-Einstein condensate, the phase collective 
mode will have a roton minimum at a wavevector inversely proportional to the 
interparticle distance.~\cite{palo2,feynman} In the vicinity of this 
minimum, for $\bk\simeq \bk_0$, the dispersion will be gapped with
$\omega_{\bk}\simeq \Delta_r+(\bk-\bk_0)^2/2m_r$ where $\Delta_r$ is the 
energy gap at the roton minimum and $m_r$ is the roton mass. We postulate, 
based on results for collective mode dispersion in similar 
systems,~\cite{palo2,qhe}, that as $d/a_B$ increases the roton gap 
$\Delta_r$ is suppressed, reflecting the tendency towards a state with a broken 
translational symmetry with characteristic wavevector $k_0=(r_sa_B)^{-1}$. 
It follows (from the existence of a ring of roton modes which are softening) 
that the system "jumps" into Wigner supersolid phase via a first-order phase 
transition, as it does in the case of liquid-to-solid transition in $^4$He under
pressure.~\cite{nozieres} Similarly, we postulate that the roton gap remains 
finite when the crystalline order emerges, as it does in $^4$He and in 
quantum Hall bilayers; the magnitude of the roton softening can only be 
addressed by a microscopic calculation.~\cite{palo2}

\noindent{\it How to detect a Wigner Supersolid?} Observation of a 
Wigner supersolid, with two nonzero order parameters, will require separate 
measurements of phase coherence and crystalline order. Let us consider how 
the observable properties change when we enter Wigner supersolid phase from 
the uniform dipolar superfluid phase or the electron-hole Wigner crystal 
phase. The uniform dipolar superfluid phase is characterized by 
dissipationless counterflow currents proportional to the in-plane magnetic 
field,~\cite{dipolar} increased exciton recombination 
rate,~\cite{butov,snoke,lai} and, because the excitons are delocalized, an 
enhanced interlayer tunneling conductance,~\cite{em} When the system becomes 
supersolid, the phase coherence will be manifest in enhanced recombination rate
and dissipationless counterflow. However, since the excitons are localized, 
interlayer tunneling conductance will be {\it suppressed} compared to its 
value in the superfluid phase. Since the exciton recombination in the Wigner 
crystal is restricted to lattice sites, {\it Fourier transform of 
spatially resolved photoluminescence will reflect the crystalline structure}. 
The uncorrelated Wigner crystal phase is characterized by insulating behavior 
for in-plane and interlayer transport, as well as phonons with dispersion
$\omega_p\propto k^{1/2}$ since the Coulomb interaction is 
$V(r)=e^2/\epsilon r$. When the two Wigner crystals become phase
coherent and the system becomes a supersolid, the in-plane transport
will support dissipationless counterflow, and the phonon dispersion will
change to $\omega_p\propto k^{3/2}$ since the dipolar interaction is
$V_d(r)=e^2d^2/\epsilon r^3$. In addition, the interlayer tunneling 
conductance will be {\it enhanced} compared to its value in the 
electron-hole Wigner crystal state, due to increased recombination rate.

Thus, the existence of a Wigner supersolid can be confirmed by transport 
measurements (interlayer tunneling conductance, counterflow superfluidity) 
and spatially resolved photoluminescence measurements.

\noindent {\it Conclusion:} We have shown, based on general principles, 
the existence of a supersolid phase in low-density electron-hole bilayers 
with moderate interlayer separations. Since the critical temperature for BEC 
in two dimensions is $T=0$, we have focused on quantum phase transitions 
between possible ground states of this system in the absence of disorder; 
thus, our discussion is only applicable at low temperatures and to ideally 
pure systems. Experimental observation of the proposed supersolid phase will 
require studying the effects of disorder and finite temperature, which are 
beyond the scope of this note. We believe that at finite temperatures and in 
the presence of finite disorder, our proposed Wigner supersolid is likely to 
be a crossover phase with the strict supersolid existing only at $T=0$. More 
theoretical work will be needed to address this question.

We have shown that experimental observation of the Wigner supersolid will 
require separate probes looking at the phase-coherence response and intralayer 
Wigner crystal response. Such a supersolid behavior should exhibit strong 
dependence on layer separation, temperature, and $r_s$ since, as we have 
shown, the supersolid phase will only be stable in the region 
$\sqrt{r_s}\leq d/a_B\leq r_s$, at low temperatures. Although the 
experimental observation of our proposed Wigner supersolid may be difficult, 
we believe that its existence in electron-hole bilayers is a robust 
conclusion. The exact values of $d/a_B$ and $r_s$ at which Wigner 
supersolid is realized will have to be determined via Monte Carlo 
simulations.~\cite{palo2} Wigner crystallization in electron-hole bilayers 
occurs~\cite{palo} at $r_s\sim 20$ compared with~\cite{tanatar} $r_s\sim 40$ 
in the single-layer system and is thus energetically favorable in bilayer
systems. This observation makes the case stronger for the existence of our 
proposed Wigner supersolid phase.

Wigner supersolid of excitons in semiconductors is a natural extension of 
the excitonic condensate, just as a supersolid is a natural extension of 
Bose-Einstein condensate with repulsive 
interactions.~\cite{gabor,palo2,scarola,hcbosons}. It will be interesting to
explore the consequences of a similar analysis in electron-electron 
(or hole-hole) bilayers where spontaneous interlayer phase coherence may 
exist~\cite{zheng} in the absence of a magnetic field, as well as in the
case of quantum Hall bilayers~\cite{demler}. The verification (or 
falsification) of our proposed Wigner supersolid will further our 
understanding of the interplay between interparticle interactions and 
quantum statistics.

It is a pleasure to thank to A.H. MacDonald, S.M. Girvin and Jinwu
Ye for useful conversations. This work was supported by US DOE
through LDRD (AVB), and NSF and ONR (SDS). AVB and SDS are grateful to
Kavli Institute for Theoretical Physics for hospitality at the
earlier stages of the work on this paper.



\begin{thebibliography}{99}
\bibitem{cav1} We assume, without any loss of generality, that the
electron-hole bilayer is described by a single $r_s$ parameter, {\it i.e.}
the electrons and holes have the same band mass. We also assume that the
exciton mass is the same as the carrier band mass. This does not change the
{\it qualitative} form of the phase boundary.
\end{thebibliography}
\end{document}